\documentclass[aps,prd,twocolumn,nofootinbib]{revtex4-1}
\usepackage{epsfig}
\usepackage[colorlinks,linkcolor=blue,anchorcolor=blue,citecolor=blue,urlcolor=blue,breaklinks=true]{hyperref}
\usepackage{graphics}
\usepackage{multirow}
\usepackage{slashed}
\usepackage{color}
\usepackage{amsmath}
\usepackage{bm}
\usepackage{setspace}
\emergencystretch=\maxdimen
\begin{document}
\author{Ya-Peng Zhao$^{1}$}\email{zhaoyapeng2013@hotmail.com}
\author{Shu-Yu Zuo$^{2}$}
\author{Cheng-Ming Li$^{3}$}\email{licm@zzu.edu.cn}
\address{$^{1}$ Collage of Physics and Electrical Engineering, Anyang Normal University, Anyang, 455000, China}
\address{$^{2}$ College of Science, Henan University of Technology, Zhengzhou 450000, China}
\address{$^{3}$ School of Physics and Microelectronics, Zhengzhou University, Zhengzhou 450001, China}
\title{QCD phase diagram and critical exponents within the nonextensive Polyakov-Nambu-Jona-Lasinio model}

\begin{abstract}
   We present a nonextensive version of the Polyakov-Nambu-Jona-Lasinio model which is based on the nonextentive statistical mechanics. This new statistics is characterized by a dimensionless nonextensivity parameter $q$ that accounts for all possible effects violating the assumptions of the Boltzmann-Gibbs (BG) statistics (when $q\rightarrow1$, it returns to the BG case). Based on the nonextensive Polyakov-Nambu-Jona-Lasinio model, we discussed the influence of the nonextensive effects on the QCD phase transition, especially on the location of the critical end point (CEP). A new and interesting phenomenon we found is that with the increase of $q$, the CEP position initially shifted toward the direction of larger chemical potential and lower temperature. But then, when $q$ is greater than a critical value $q_{c}$, the CEP position moves in the opposite direction. In other words, as $q$ increases, the CEP position moves in the direction of smaller chemical potential and higher temperature. In addition, we calculated the influence of the nonextensive effects on the critical exponents and found that they remain almost constant with $q$.
\bigskip

\bigskip

\noindent PACS Number(s): 12.38.Mh, 12.39.-x, 25.75.Nq, 12.38.Aw

\end{abstract}

% QED3 11.10.Kk, 11.15.Tk, 11.30.Qc
% quark-gluon, 12.38.Mh; Quark models, 12.39.-x; Quark confinement, 12.38.Aw; Quark deconfinement, 25.75.Nq; Lattice QCD calculations, 12.38.Gc;
% Quark-gluon plasma, 12.38.Mh; phase transitions in QGP, 25.75.Nq; production of QGP, 25.75.Nq; particle physics of chirality, 11.30.Rd;

\maketitle

\section{INTRODUCTION}
The QCD phase diagram, especially its critical end point (CEP), is one of the most important aspects of strongly interacting matter. From the experimental side,
the Beam Energy Scan (BES) program at the Relativistic Heavy-Ion Collider (RHIC) facility aims to identify signals of the expected CEP. While the goal of the CBM experiment at the future Facility for Antiproton and Ion Research (FAIR) at GSI is to explore the QCD phase diagram in the region of high baryon densities and search for the expected first-order phase transition.
From the theoretical side, people use various methods to study the QCD phase diagram. Such as chiral perturbation theory~\cite{Espriu:2020dge}, finite energy sum rules~\cite{ PhysRevD.84.056004}, Dyson-Schwinger equations~\cite{Fischer:2018sdj,PhysRevD.91.056003,Zhao_2019}, Nambu-Jona-Lasinio (NJL) model and
Polyakov-Nambu-Jona-Lasinio (PNJL)~\cite{BUBALLA2005205,Cui:2018bor,PhysRevC.101.065203,PhysRevD.101.096006,ZHAO2020114919}.

However, it is worth noting that when people study the QCD phase transition, a statistical method often used is Boltzmann-Gibbs (BG) statistics.
Strictly speaking, the BG statistics can only be applied to systems in equilibrium and within the thermodynamic limit. Obviously, in the relativistic heavy-ion collisions, in which the quark-gluon plasma (QGP) produced experiences strong intrinsic fluctuations and long-range correlations. The volume of QGP is small enough and it evolves rapidly. Therefore, this system is far from being uniform and no global equilibrium is established. As a result, some quantities become nonextensive and develop power-law tailed rather than exponential distributions. In such cases the application of the usual BG statistics is questionable.

Thus, a nonextensive statistics was first proposed by Tsallis, also known as Tsallis statistics~\cite{Tsallis1988}. The most typical feature of Tsallis statistics is that it replaces the usual exponential factors by their q-exponential equivalents~\cite{PhysRevC.77.044903,Eur.Phys.J.A2016,shen2017chiral},
\begin{eqnarray}\label{Tsallis}
P_{BG}(E)=exp(-\frac{E}{T})\longrightarrow P_{q}(E)=exp_{q}(-\frac{E}{T}),
\end{eqnarray}
where
\begin{eqnarray}
exp_{q}(x)=[1+(1-q)x]^{\frac{1}{1-q}},
\end{eqnarray}
correspondingly, its inverse function is
\begin{eqnarray}
ln_{q}(x)=\frac{x^{1-q}-1}{1-q}.
\end{eqnarray}
The non-extensivity parameter $q$ represents all possible factors that do not satisfy the BG statistical assumptions.
Its physical interpretation is currently not very clear. For $q>1$, the most popular one is that $q-1$ measures the intrinsic temperature fluctuations of the system considered~\cite{PhysRevLett.94.132302,PhysRevLett.84.2770}.
For $q<1$, it is usually attributed to some specific correlations limiting the available phase space~\cite{Kodama:2008br} or to the possible fractality of the allowed phase space~\cite{GarciaMorales:2005in}. Here we do not discuss the meaning of $q$ but regard $q-1$ as a description of deviations from BG statistics.
When $q\rightarrow1$, $exp_{q}(x)\rightarrow exp(x)$, $ln_{q}(x)\rightarrow ln(x)$ and Tsallis statistics returns to BG statistics.

Tsallis statistics have been applied to many branches of physics.  such as high-energy physics~\cite{BEDIAGA2000156,Wilk2012,LI2013352,PhysRevD.91.054025,De2014,Bhattacharyya2016,PhysRevC.83.064903,PhysRevC.75.064901,Aamodt2011,Aad:2010ac,Khachatryan2010}, astrophysics~\cite{Tsallis:2012js,Lavagno:2011zm}, cold atoms in optical lattices~\cite{PhysRevLett.96.110601}, anomalous diffusion~\cite{PhysRevLett.115.238301,PhysRevE.85.021146}, among many others.
In particular, Refs~\cite{tirnakli2016the,cirto2018validity} show us in a very clear way that with the ergodicity breakdown, the failure of BG statistics and the emergence of Tsallis statistics. And Ref.~\cite{PhysRevLett.115.238301} experimentally validate a particular case of the nonextensive scaling law in confined granular media.
In addition, it is interesting that Ref~\cite{Javidan:2020lup} studies the nonextensive effects of the QCD running coupling constant $\alpha_{s}$, and successfully dealt with the inconsistency between the theory and experiment in the non-perturbation region. Finally, more about Tsallis' statistics and its diverse applications can be found in Ref.~\cite{tsallis2009introduction}.

As mentioned above, in order to be as consistent as possible with the real experimental environment in which the QCD phase transition occurs, using Tsallis statistics is a better choice. Therefore, we generalize the PNJL model to its nonextensive version. Compared with NJL model, this model has proven to be more successful in reproducing lattice data concerning QCD thermodynamics~\cite{PhysRevD.73.014019}. Besides, other models such as the linear sigma model and NJL model have also been generalized to its nonextensive version to study the thermodynamic quantities of the QCD matter and its phase diagram~\cite{Eur.Phys.J.A2016,shen2017chiral}.

This paper is organized as follows: In Sec. II, we introduce the nonextensive version of the PNJL model.
In Sec. III, we discuss the influence of nonextensive effects on the QCD phase transition, especially the position of CEP, and the critical exponents. Finally, we give a brief summary of our work in Sec. IV.
\section{PNJL and nonextensive pnjl model}\label{ok}
\subsection{PNJL model}
Before introducing the nonextensive PNJL model, let's make a basic introduction to the PNJL model.
The Lagrangian of the two-flavor and three-color PNJL model reads~\cite{PhysRevD.73.014019}
\begin{eqnarray}
\mathcal{L}_{PNJL} &=& \bar{\Psi}(i\gamma_{\mu}D^{\mu}-\hat{m})\Psi +G\,[(\bar{\Psi}\Psi)^2+(\bar{\Psi}i\gamma_5\vec{\tau}\Psi)^2]\nonumber\\ &&-\mathcal{U}(\Phi,\bar{\Phi};T),
\label{effective}
\end{eqnarray}
where $\Psi=(u,d)$ represents the two flavor quark field with three colors and $\hat{m}=diag(m_{u},m_{d})$ with $m_{\mu}=m_{d}=m$ stands for the current quark mass matrix. $\tau^{a}(a=1,2,3)$ corresponds to the Pauli matrices in flavor space and $G$ is the effective coupling strength of four point interaction of quark fields.

The effective Polyakov-loop potential $\mathcal{U}(\Phi,\bar{\Phi};T)$ is expressed in terms of the traced Polyakov-Loop expectation value $\Phi$ and its conjugate
\begin{eqnarray}
\Phi=\frac{\langle Tr_{c}L \rangle}{N_{c}},     \bar{\Phi}=\frac{\langle Tr_{c}L^{\dagger} \rangle}{N_{c}}.
\end{eqnarray}
The Polyakov-loop $L$ is defined as
\begin{eqnarray}
L(\vec{x})=P\exp(i\int_{0}^{\beta}A_{4}(\vec{x},\tau)d\tau),
\end{eqnarray}
where $A_{4}=iA_{0}$ is the temporal component of Euclidian gauge field $(\vec{A},A_{4})$, $\beta=1/T$, and $P$ denotes the path ordering. Morever, for simplicity we take the approximation $L^{\dag}=L$ following Refs.~\cite{PhysRevD.84.014011,PhysRevD.85.054013,PhysRevD.94.071503,PhysRevD.102.014014}, which implies $A_{4}^{8}=0$.

The thermodynamic potential density function can be
determined in the mean field approximation as:
\begin{eqnarray}
\Omega(\mu,T,M,\Phi,\bar{\Phi})&=&\mathcal{U}(\Phi,\bar{\Phi};T)+\frac{(M-m)^{2}}{4G}\\
&-&2N_{c}N_{f}\int_{0}^{\Lambda}\frac{{\rm d}^3\vec{p}}{(2\pi)^3}E_{p}\nonumber\\
&-&2N_{f}T\int_{0}^{\infty}\frac{{\rm d}^3\vec{p}}{(2\pi)^3}(lnF^{+}+lnF^{-}),\nonumber
\end{eqnarray}
where $M$ means the dynamical quark mass. It relates to the quark chiral condensate $\sigma=\langle\bar{\Psi}\Psi\rangle$ as follows
\begin{eqnarray}
M=m-2G\sigma,
\end{eqnarray}
and
\begin{eqnarray}
F^{+}&=&1+3(\Phi+\bar{\Phi}e^{-\frac{E_{p}-\mu}{T}})e^{-\frac{E_{p}-\mu}{T}}+e^{-3\frac{E_{p}-\mu}{T}},\nonumber\\
F^{-}&=&1+3(\bar{\Phi}+\Phi e^{-\frac{E_{p}+\mu}{T}})e^{-\frac{E_{p}+\mu}{T}}+e^{-3\frac{E_{p}+\mu}{T}},
\end{eqnarray}
in which $E_{p}=\sqrt{p^{2}+M^{2}}$ is the single quasi-particle energy. In the above integrals, following Refs.~\cite{sym2031338,PhysRevD.73.014019,PhysRevC.79.055208,FUKUSHIMA2004277}, the vacuum integral has a cut-off $\Lambda$ whereas the medium dependent integrals have been extended to infinity.

The two Polyakov-loop effective potentials used in this paper are as follows:

(1) The polynomial effective Polyakov-loop potential is~\cite{PhysRevD.62.111501,PhysRevC.66.034903,PhysRevD.73.014019}
\begin{eqnarray}
\frac{\mathcal{U_{P}}}{T^{4}}=-\frac{b_{2}(T)}{2}\bar{\Phi}\Phi-\frac{b_{3}}{6}(\Phi^{3}+\bar{\Phi}^{3})+\frac{b_{4}}{4}(\bar{\Phi}\Phi)^{2},
\end{eqnarray}
with a temperature-dependent coefficient
\begin{eqnarray}
b_{2}(T)=a_{0}+a_{1}(\frac{T_{0}}{T})+a_{2}(\frac{T_{0}}{T})^{2}+a_{3}(\frac{T_{0}}{T})^{3},
\end{eqnarray}
and the corresponding parameters are given in Table~\ref{tb1}.
\begin{center}
\begin{table}
\caption{Parameter set used in our work.}\label{tb1}
\begin{tabular}{p{1.2cm} p{1.2cm} p{1.2cm} p{1.2cm} p{1.2cm} p{1.2cm}}
%\begin{tabular}{cccccc}
\hline\hline
$a_0$&$a_1$&$a_2$&$a_3$&$b_3$&$b_4$\\
\hline
6.75&-1.95&2.625&-7.44&0.75&7.5\\
\hline\hline
\end{tabular}
\end{table}
\end{center}

(2) The Logarithmic effective Polyakov-loop potential is~\cite{PhysRevD.75.034007}
\begin{eqnarray}
\frac{\mathcal{U_{L}}}{T^{4}}&=&-\frac{a(T)}{2}\Phi\bar{\Phi}+b(T)ln[1-6\Phi\bar{\Phi}-3(\Phi\bar{\Phi})^{2}\nonumber\\
&&+4(\Phi^{3}+\bar{\Phi}^{3})],
\end{eqnarray}
with the temperature-dependent coefficients
\begin{eqnarray}
a(T)=a_{0}+a_{1}(\frac{T_{0}}{T})+a_{2}(\frac{T_{0}}{T})^{2},
\end{eqnarray}
and
\begin{eqnarray}
b_{T}=b_{3}(\frac{T_{0}}{T})^{3},
\end{eqnarray}
the corresponding parameters are given in Table~\ref{tb2}. Here, the logarithmic form constrains $\Phi,\bar{\Phi}\leq1$.
\begin{center}
\begin{table}
\caption{Parameter set used in our work.}\label{tb2}
\begin{tabular}{p{1.8cm} p{1.8cm} p{1.8cm} p{1.8cm}}
%\begin{tabular}{cccccc}
\hline\hline
$a_0$&$a_1$&$a_2$&$b_3$\\
\hline
3.51&-2.47&15.2&-1.75\\
\hline\hline
\end{tabular}
\end{table}
\end{center}

\begin{center}
\begin{table}
\caption{Parameter set used in our work.}\label{tbnjl}
\begin{tabular}{p{2.4cm} p{2.4cm} p{2.4cm}}
%\begin{tabular}{cccccc}
\hline\hline
$\Lambda(\mathrm{MeV})$&$G(\mathrm{MeV^{-2}})$&$m(\mathrm{MeV})$\\
\hline
651&$5.04\times10^{-6}$&5.5\\
\hline\hline
\end{tabular}
\end{table}
\end{center}

In a pure gauge sector, $T_{0}=270\ \mathrm{MeV}$. However, in the presence of dynamical quarks, the critical temperature $T_{0}$ will have an $N_{f}$ dependence $T_{0}(N_{f})$. Here, we let $T_{0}(2)=192\ \mathrm{MeV}$  follows Ref.~\cite{PhysRevD.76.074023}.
Besides, the parameters for the NJL model part of the effective Lagrangian $\mathcal{L}_{PNJL}$ are summarized in Table~\ref{tbnjl}. The resulting physical quantities are $f_{\pi}=92.3\ \mathrm{MeV}$, $m_{\pi}=139.3\ \mathrm{MeV}$ and $-\langle\bar{\Psi}_{u}\Psi_{u}\rangle^{\frac{1}{3}}=251\ \mathrm{MeV}$~\cite{PhysRevD.73.014019}.

Finally, the solutions of the mean field equations are obtained by minimizing the thermodynamic potential function $\Omega$ with respect to $M$ and $\Phi$, that is
\begin{eqnarray}\label{gap}
\frac{\partial\Omega}{\partial M}=\frac{\partial\Omega}{\partial\Phi}=0.
\end{eqnarray}
\subsection{nonextensive PNJL model}
In short, when we use Tsallis statistics instead of BG statistics to describe a system, it means we need to do the replacement as shown in Eq.~(\ref{Tsallis}). And we will take up two simplifications in the following calculations as in Ref.~\cite{PhysRevD.101.096006}.

(i) The non-extensive effects are not considered in the pure Yang-Mills sector. that is to say, the Polyakov-loop potential remains unchanged and feels nonextensive effects implicitly only through the saddle point equations.

(ii) The usual PNJL model parameters remain unchanged.
We treat $q$, just as people treat volume $V$ in the study of finite-size effects, as a thermodynamic variable in the same footing as $T$ and $\mu$~\cite{PhysRevD.87.054009,Grunfeld2018}. In fact, this is all based on the ansatz that the parameters determined at zero temperature and zero quark chemical potential can be used to study the finite temperature and finite quark chemical potential. It is also pointed out in the Refs.~\cite{PhysRevD.82.076003,Cui2014} that the coupling constant $G$ should be dependent on the temperature and the quark chemical potential by depending on the order parameter $\Phi$ or $\langle\bar{\Psi}\Psi\rangle$. But here, we do not consider this situation.

Thus, within the q-PNJL model, the thermodynamic potential density function becomes
\begin{eqnarray}\label{omega00}
\Omega_{q}(\mu,T,M,\Phi,\bar{\Phi})&=&\mathcal{U}(\Phi,\bar{\Phi};T)+\frac{(M-m)^{2}}{4G}\\
&-&2N_{c}N_{f}\int_{0}^{\Lambda}\frac{{\rm d}^3\vec{p}}{(2\pi)^3}E_{p}\nonumber\\
&-&2N_{f}T\int_{0}^{\infty}\frac{{\rm d}^3\vec{p}}{(2\pi)^3}(ln_{q}F_{q}^{+}+ln_{q}F_{q}^{-}),\nonumber
\end{eqnarray}
where
\begin{eqnarray}
F_{q}^{+}=&1&+3(\Phi+\bar{\Phi}e_{q}(-\frac{E_{p}-\mu}{T}))e_{q}(-\frac{E_{p}-\mu}{T})\nonumber\\
&+&e_{q}(\frac{-3(E_{p}-\mu)}{T}),\nonumber\\
F_{q}^{-}=&1&+3(\bar{\Phi}+\Phi e_{q}(-\frac{E_{p}+\mu}{T}))e_{q}(-\frac{E_{p}+\mu}{T})\nonumber\\
&+&e_{q}(\frac{-3(E_{p}+\mu)}{T}).
\end{eqnarray}
In this paper we consider only $q>1$ because of the typical value of the non-extensivity parameter $q$ for high energy collisions is found to be $1 \leq q \leq 1.2$~\cite{CLEYMANS2013351,LI2013352,PhysRevD.91.054025,Azmi_2014}.
In order to ensure that $e_{q}(x)$ is always a non-negative real function, the following condition must be supplemented (known as Tsallis cut-off prescription)
\begin{eqnarray}
e_{q}(x)=0, \quad \text{for} \quad [1+(1-q)x]<0,
\end{eqnarray}

Besides, it is important to realize that for $T\rightarrow0$ one always gets $\Omega_{q}\rightarrow\Omega$ as long as $q>1$. This means that we can expect any nonextensive signature only for high enough temperatures.

For studying the QCD phase diagram, according to Eq.~(\ref{gap}), we need to solve the following coupled equations:
\begin{widetext}
\begin{eqnarray}
M&=&m+4GN_{c}N_{f}\int\frac{{\rm d}^3\vec{p}}{(2\pi)^3}\frac{M}{E_{p}}[1-n_{q}-\bar{n}_{q}],\\
0&=&\frac{\partial\mathcal{U}}{\partial\Phi}-2N_{c}N_{f}T\int_{0}^{\infty}\frac{{\rm d}^3\vec{p}}{(2\pi)^3}\{\frac{(1+e_{q}(\frac{-(E_{p}-\mu)}{T}))e_{q}(\frac{-(E_{p}-\mu)}{T})}{[1+3\Phi(1+e_{q}(\frac{-(E_{p}-\mu)}{T}))e_{q}(\frac{-(E_{p}-\mu)}{T})+e_{q}(\frac{-3(E_{p}-\mu)}{T})]^{q}}\nonumber\\
&&+\frac{(1+e_{q}(\frac{-(E_{p}+\mu)}{T}))e_{q}(\frac{-(E_{p}+\mu)}{T})}{[1+3\Phi(1+e_{q}(\frac{-(E_{p}+\mu)}{T}))e_{q}(\frac{-(E_{p}+\mu)}{T})+e_{q}(\frac{-3(E_{p}+\mu)}{T})]^{q}}\},
\end{eqnarray}
where the q-version of the Fermi-Dirac distribution is
\begin{eqnarray}\label{nq}
n_{q}(T,\mu)=\frac{e_{q}^{q}(\frac{-3(E_{p}-\mu)}{T})+\Phi(1+2e_{q}(\frac{-(E_{p}-\mu)}{T}))e_{q}^{q}(\frac{-(E_{p}-\mu)}{T})}{[1+3\Phi(1+e_{q}(\frac{-(E_{p}-\mu)}{T}))e_{q}(\frac{-(E_{p}-\mu)}{T})+e_{q}(\frac{-3(E_{p}-\mu)}{T})]^{q}},\nonumber\\
\end{eqnarray}
and
\begin{eqnarray}\label{nbarq}
\bar{n}_{q}(T,\mu)=\frac{e_{q}^{q}(\frac{-3(E_{p}+\mu)}{T})+\Phi(1+2e_{q}(\frac{-(E_{p}+\mu)}{T}))e_{q}^{q}(\frac{-(E_{p}+\mu)}{T})}{[1+3\Phi(1+e_{q}(\frac{-(E_{p}+\mu)}{T}))e_{q}(\frac{-(E_{p}+\mu)}{T})+e_{q}(\frac{-3(E_{p}+\mu)}{T})]^{q}}.\nonumber\\
\end{eqnarray}
\end{widetext}
When $q\rightarrow1$, they return to the distribution function of the usual PNJL model.

\begin{figure}
\includegraphics[width=0.47\textwidth]{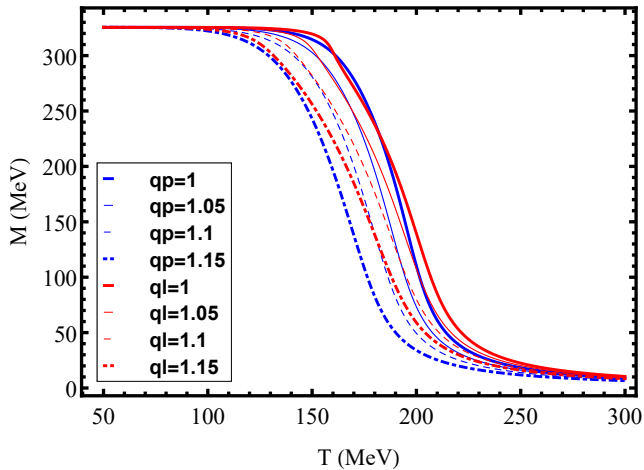}
\caption{Constituent quark mass $M$ as a function of $T$ at $\mu=0$ for two different potentials $\mathcal{U}$ and four parameters $q$. Where $\mathrm{l}$, and $\mathrm{p}$ represent logarithmic and polynomial Polyakov-loop potential, respectively.}
\label{Fig:M}
\end{figure}
\begin{figure}
\includegraphics[width=0.47\textwidth]{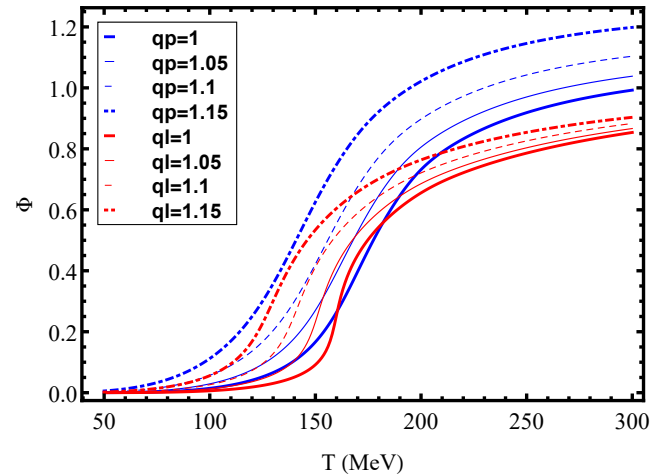}
\caption{Polyakov-loop expectation value $\Phi$ as a function of $T$ at $\mu=0$ for two different Polyakov-loop potentials $\mathcal{U}$ and four parameters $q$.}
\label{Fig:phi}
\end{figure}
\section{QCD phase transition and critical exponents within tsallis statistics}\label{bigtwo}
\subsection{QCD phase transition}
As a first step, we plot $M$ and $\Phi$ as a function of $T$ for four different $q$, ($q=1$, $1.05$, $1.1$, $1.15$) as well as two different $\mathcal{U}$ ($\mathcal{U_{P}}$, $\mathcal{U_{L}}$) as shown in Figs.~\ref{Fig:M},~\ref{Fig:phi}. We found that the nonextensivity parameter $q$ does not change the conclusion that the finite-temperature QCD transition is not a real phase transition, but a crossover~\cite{Aoki:2006we}. However, as $q$ increases, the transition occurs at a smaller pseudo-critical temperature $T_{c}$. The same conclusion also appears in the non-extensive linear sigma model~\cite{shen2017chiral}. In addition, it should be noted that $q$ does not have any effect on the QCD phase transition at zero temperature.

Next, we are more concerned about the impact of the nonextensive effects on the CEP position. As we know, in the neighborhood of the CEP position, the susceptibility tends to diverge. Therefore, we can determine the position of CEP by the following thermal susceptibility
\begin{figure}
\includegraphics[width=0.47\textwidth]{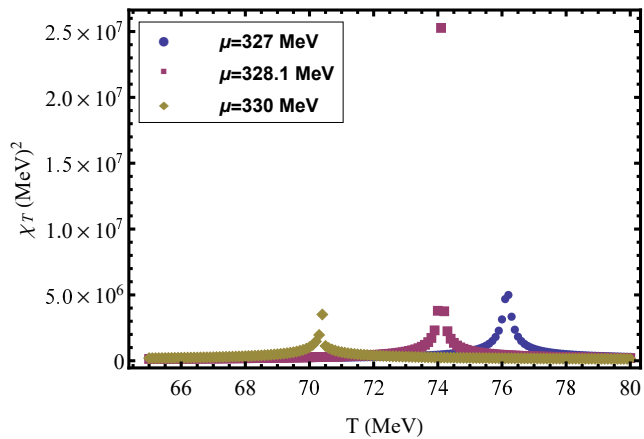}
\caption{The susceptibility $\chi_{T}$ as a function of $T$ at $q=1$ for Polyakov-loop potential $\mathcal{U_{P}}$ and three different quark chemical potentials.}
\label{1}
\end{figure}
\begin{figure}
\includegraphics[width=0.47\textwidth]{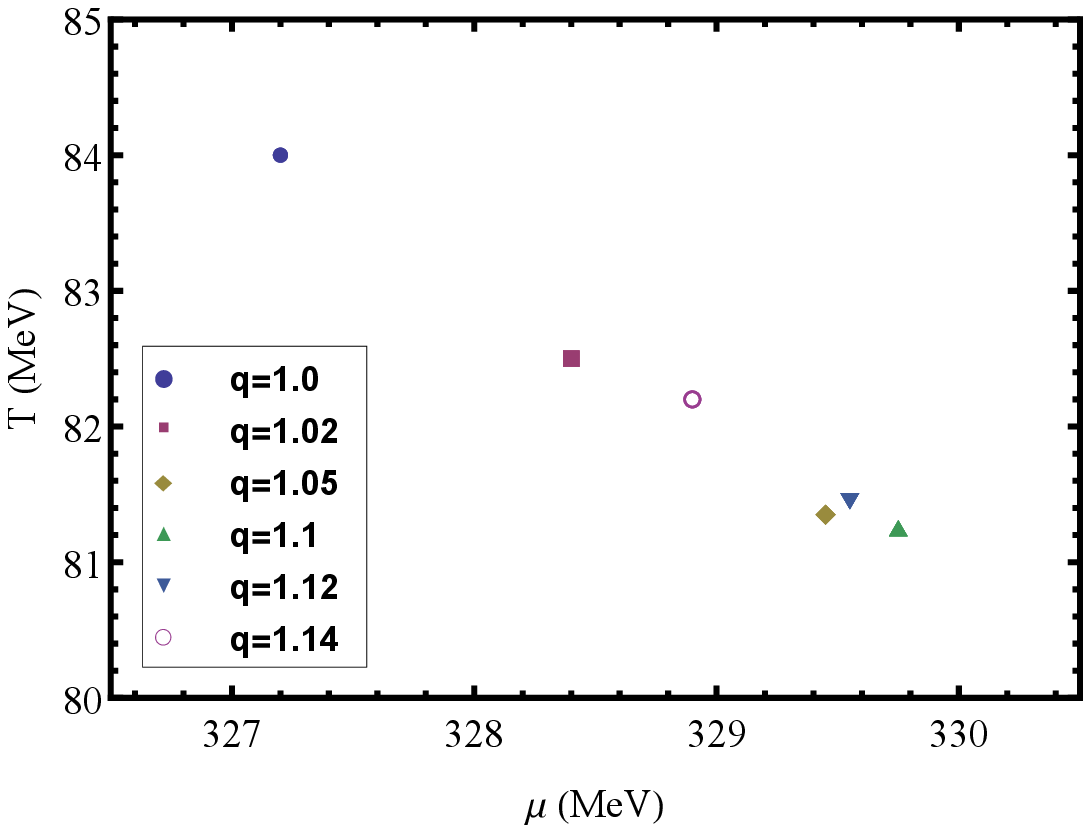}
\caption{The trajectory of CEP position with $q$ in the $T$-$\mu$ plane for Polyakov-loop potential $\mathcal{U_{L}}$.}
\label{2}
\end{figure}
\begin{figure}
\includegraphics[width=0.47\textwidth]{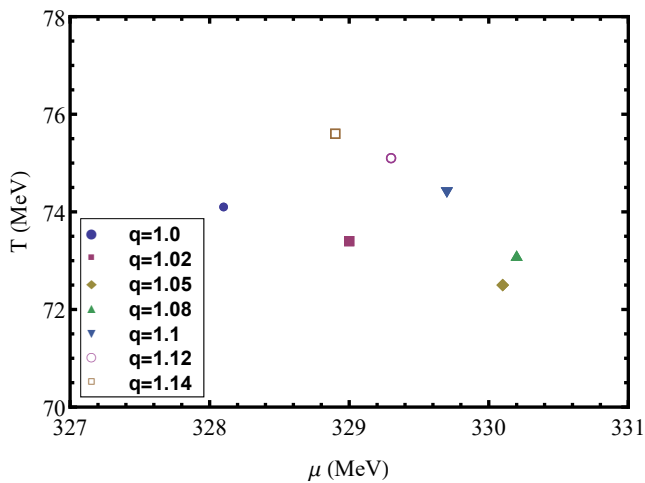}
\caption{The trajectory of CEP position with $q$ in the $T$-$\mu$ plane for Polyakov-loop potential $\mathcal{U_{P}}$.}
\label{3}
\end{figure}
\begin{eqnarray}
\chi_{T}=\frac{\partial\sigma}{\partial T}.
\end{eqnarray}
From Fig.~\ref{1}, we can clearly see that when $\mu=330\ \mathrm{MeV}$, the susceptibility is discontinuous and corresponds to a first-order phase transition. When $\mu=328.1\ \mathrm{MeV}$, the susceptibility tends to diverge, corresponding to the position of CEP and the susceptibility is continuous when $\mu=327\ \mathrm{MeV}$, corresponding to a crossover transition.
The effect of nonextensivity parameter $q$ on the position of CEP is shown in Figs~\ref{2},~\ref{3}. The most interesting phenomenon we found is that with the increase of $q$, at the beginning, the position of CEP moves toward the direction of larger chemical potential and lower temperature. But then, when $q$ is greater than a critical value $q_{c}$, the CEP position moves in the opposite direction. In other words, as $q$ increases, CEP moves in the direction of smaller chemical potential and higher temperature. Obviously, this interesting reentry phenomenon is independent of the choice of Polyakov-loop potentials. And for $\mathcal{U_{L}}$, $\mathcal{U_{P}}$, the critical values $q_{c}$ are $1.1$ and $1.08$, respectively.
\subsection{critical exponents}
\begin{figure}
\includegraphics[width=0.47\textwidth]{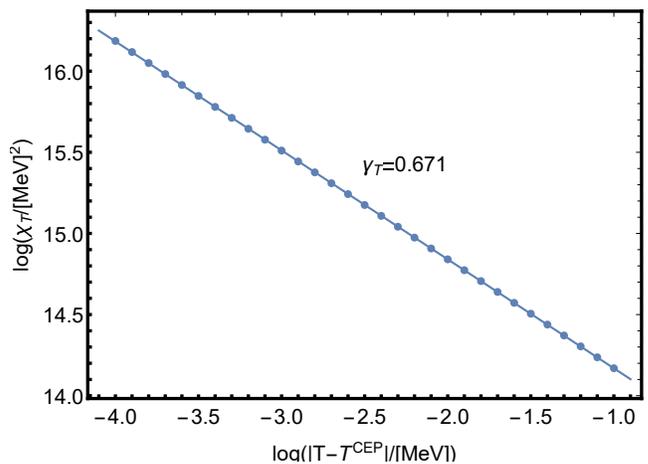}
\caption{The logarithm value of the thermal susceptibility $\chi_{T}$ as a
function of log$|T-T^{\mathrm{CEP}}|$ at the fixed quark chemical potential $\mu^{\mathrm{CEP}}$ for Polyakov-loop potential $\mathcal{U_{P}}$.}
\label{ce}
\end{figure}
\begin{figure}
\includegraphics[width=0.47\textwidth]{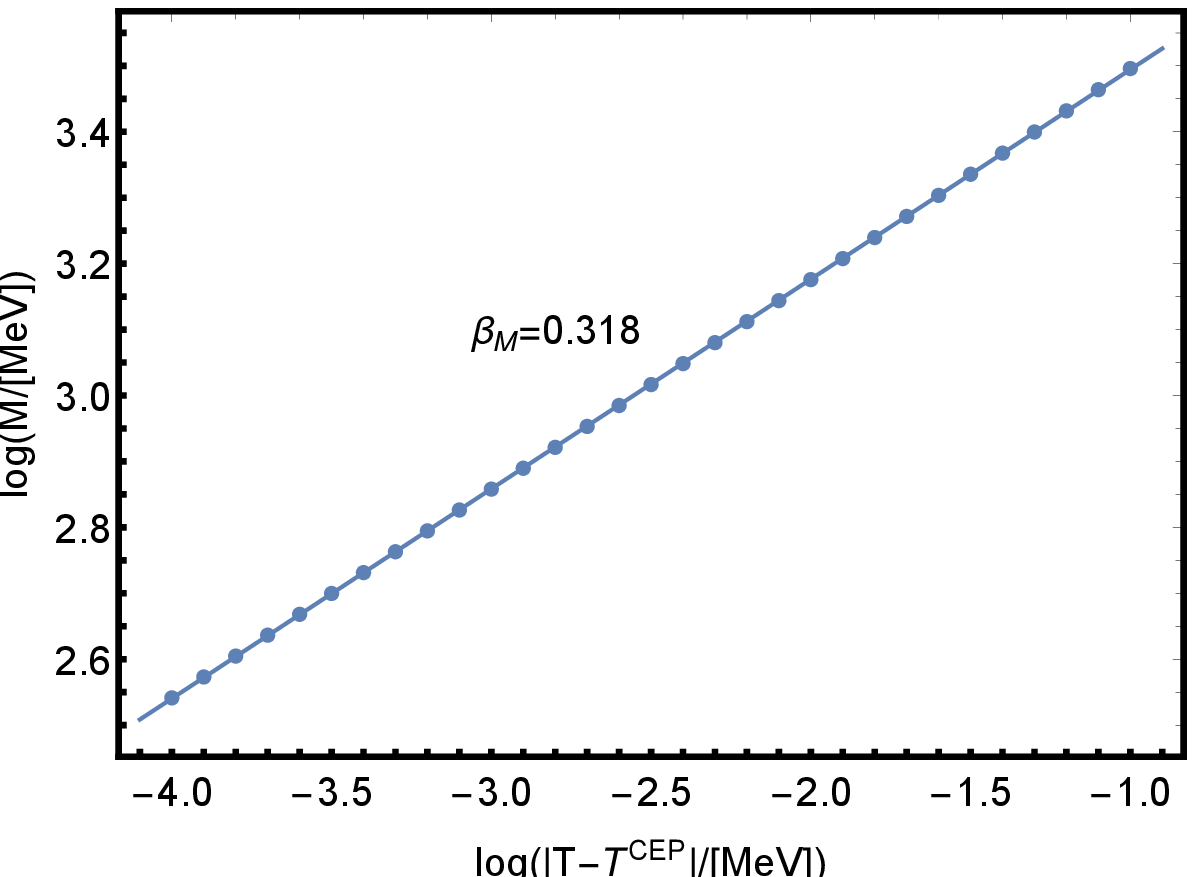}
\caption{The logarithm value of the quark mass $M$ as a
function of log$|T-T^{\mathrm{CEP}}|$ at the fixed quark chemical potential $\mu^{\mathrm{CEP}}$ for Polyakov-loop potential $\mathcal{U_{P}}$.}
\label{cem}
\end{figure}
\begin{center}
\begin{table}
\caption{The dependence of the critical exponents on the nonextensivity parameter $q$ for Polyakov-loop potential $\mathcal{U_{L}}$.}\label{tb5}
\begin{tabular}{p{1.8cm} p{2.2cm} p{1.5cm} p{2.5cm}}
%\begin{tabular}{cccccc}
\hline\hline
$\mathrm{q}$&$\mathrm{Quantity}$&$\mathrm{Path}$&$\mathrm{Numerical}\ \mathrm{result}$\\
\hline
\multirow{4}*{$q=1.0$}& \multirow{2}*{$\gamma_{T}$}&$\uparrow$
&0.677\\
& &$\downarrow$&0.678\\
& \multirow{2}*{$\beta_{M}$}&$\uparrow$
&0.361\\
& &$\downarrow$&0.309\\
\hline
\multirow{4}*{$q=1.05$}& \multirow{2}*{$\gamma_{T}$}&$\uparrow$
&0.680\\
& &$\downarrow$&0.685\\
& \multirow{2}*{$\beta_{M}$}&$\uparrow$
&0.332\\
& &$\downarrow$&0.346\\
\hline
\multirow{4}*{$q=1.1$}& \multirow{2}*{$\gamma_{T}$}&$\uparrow$
&0.681\\
& &$\downarrow$&0.679\\
& \multirow{2}*{$\beta_{M}$}&$\uparrow$
&0.315\\
& &$\downarrow$&0.362\\
\hline\hline
\end{tabular}
\end{table}
\end{center}
\begin{center}
\begin{table}
\caption{The dependence of the critical exponents on the nonextensivity parameter $q$ for Polyakov-loop potential $\mathcal{U_{P}}$.}\label{tb6}
\begin{tabular}{p{1.8cm} p{2.2cm} p{1.5cm} p{2.5cm}}
%\begin{tabular}{cccccc}
\hline\hline
$\mathrm{q}$&$\mathrm{Quantity}$&$\mathrm{Path}$&$\mathrm{Numerical}\ \mathrm{result}$\\
\hline
\multirow{4}*{$q=1.0$}& \multirow{2}*{$\gamma_{T}$}&$\uparrow$
&0.671\\
& &$\downarrow$&0.678\\
& \multirow{2}*{$\beta_{M}$}&$\uparrow$
&0.318\\
& &$\downarrow$&0.353\\
\hline
\multirow{4}*{$q=1.05$}& \multirow{2}*{$\gamma_{T}$}&$\uparrow$
&0.668\\
& &$\downarrow$&0.683\\
& \multirow{2}*{$\beta_{M}$}&$\uparrow$
&0.364\\
& &$\downarrow$&0.307\\
\hline
\multirow{4}*{$q=1.1$}& \multirow{2}*{$\gamma_{T}$}&$\uparrow$
&0.660\\
& &$\downarrow$&0.673\\
& \multirow{2}*{$\beta_{M}$}&$\uparrow$
&0.370\\
& &$\downarrow$&0.304\\
\hline\hline
\end{tabular}
\end{table}
\end{center}

As we all know, in the vicinity of CEP, the divergence of susceptibility can be described by the critical exponents. Regarding the critical exponents, there are two important physical concepts. First, the scale hypothesis. The basic idea is that when approaching the critical point, the correlation length $\xi\rightarrow\infty$. And the singularity of $\xi$ determines the singularity of all thermodynamic functions. From this, the scaling law that should be satisfied between the critical indices can be derived.
Second, the universality assumption. It refers to a system with the same spatial dimension $d$ and order parameter dimension $n$, with the same critical exponent, and belonging to the same universal category. However, it should be pointed out that the validity of these concepts is based on the equilibrium phase transition system described by BG statistics. Therefore, in a system that deviates from the description of BG statistics, the critical exponents may not be completely determined by $d$ and $n$, the scaling law may  need to be reconstructed or modified~\cite{PhysRevE.80.051101,PhysRevE.102.012116,PhysRevD.97.105006,PhysRevB.93.201106,hurtado2016a}. Based on this, in this chapter we use Tsallis statistics to study the critical exponents and discuss the influence of the nonextensivity parameter $q$ on them.

Here, we choose a specific direction, which is denoted by $\uparrow$ ($\downarrow$), to calculate the critical exponents by the path from lower (higher) $T$ toward $T^{\mathrm{C}}$ (represents $T^{\mathrm{CEP}}$) with the quark chemical potential fixed at $\mu^{\mathrm{C}}$ (represents $\mu^{\mathrm{CEP}}$).
Using the linear logarithmic fit we obtain
\begin{eqnarray}
\mathrm{ln}\chi=-\gamma\mathrm{ln}|T-T^{C}|+c_{1}, \\
\mathrm{ln}|O-O^{C}|=\beta_{O}\mathrm{ln}|T-T^{C}|+c_{2},
\end{eqnarray}
$\gamma$ is the critical exponent of
susceptibility while $\beta$ is the critical exponent of order
parameter $O$, and $c_{1}$, $c_{2}$ are constants. At $q=1$ and in the direction $\uparrow$, the
fitting procedure of the critical exponents for thermal
susceptibility and quark mass is shown in Figs~\ref{ce},~\ref{cem}.

The variation of critical exponents with $q$ is shown in Tables~\ref{tb5},~\ref{tb6}. We find that when $q$ increases from $1.0$ to $1.1$, the critical exponent $\gamma_{T}$ remains almost unchanged, regardless of the Polyakov-loop potentials selected.
But for the critical exponent $\beta$, take the Polyakov-loop potential $\mathcal{U_{L}}$ as an example. We find that for the direction $\uparrow$ ($\downarrow$), $\beta_{M}$ decreases (increases) with the increase of $q$. For the Polyakov-loop potential $\mathcal{U_{P}}$, this trend is just the opposite. However, if we take the average value $\bar{\beta}_{M}=(\beta_{M\uparrow}+\beta_{M\downarrow})/2$ as the critical exponent parallel to the $T$ axis, we find that $\bar{\beta}_{M}$ is stable around $0.337$ and hardly changes with $q$.
It is worth noting that Ref.~\cite{PhysRevE.102.012116} studied the critical behavior of the two-dimensional Ising model with nonextensive statistics and found that for $q<1$, the critical exponents are related to $q$. In particular, the critical exponent $\nu$ changes with $q$ by a linear law.
\section{Summary and Conclusion}\label{summary}
In this paper, combined with the Tsallis statistics and the PNJL model, we investigated the sensitivity of the QCD phase transition and critical exponents to deviations from usual BG statistics.
Regarding the QCD phase diagram, we found that the influence of the nonextensive effects on the CEP position shows a very interesting reentry phenomenon. At the beginning, with the increase of $q$, the CEP position moves toward the direction of greater chemical potential and lower temperature. However, when $q$ is greater than the critical value $q_{c}$, as the $q$ increases, the CEP position moves in the opposite direction, that is, the direction with a smaller chemical potential and a higher temperature. Because of this reentry phenomenon, based on our calculations, we found that the nonextensive effects does not influence the CEP position as much as expected. Therefore, it may be safely ignored in the search of CEP by RHIC.
Regarding the critical exponents, numerical results based on Tsallis statistics show that the critical exponents remain almost constant with $q$. In other words, for $q>1$, it seems that the critical exponents does not depend on BG statistics or Tsallis statistics. However, this requires a more detailed argument. In addition, quark stars, as candidates for observed massive stars ($\geq2M_{\odot}$), have attracted much attention in astronomy~\cite{Li:2019ztm,Chu:2019ipr,Li:2019akk,Chen:2016ran}. Therefore, studying the influence of nonextensive effects on the structure and evolution of protoquark stars will be a very meaningful topic. These issues are what we will study in the future.

\acknowledgments
This work is supported by the Project funded by China Postdoctoral Science Foundation (Grant No. 2020M672255 and No. 2020TQ0287).
\bibliography{EPJC}
\end{document}